\documentclass[sigconf]{acmart}
\AtBeginDocument{%
  }

\setcopyright{acmlicensed}
\copyrightyear{2018}
\acmYear{2018}
\acmDOI{XXXXXXX.XXXXXXX}

\acmConference[Conference acronym 'XX]{Make sure to enter the correct
  conference title from your rights confirmation emai}{June 03--05,
  2018}{Woodstock, NY}
\acmISBN{978-1-4503-XXXX-X/18/06}

\usepackage{CJKutf8}
\usepackage{caption}
\usepackage{subfigure}

\begin{document}
\begin{CJK}{UTF8}{bsmi}

\title{Online Social Network Data-Driven Early Detection on Short-Form Video Addiction}

\author{Fang-Yu, Kuo}
\email{clairekuo0217@gmail.com}
\affiliation{%
  \institution{Department of Computer Science
National Tsing Hua University}
  \city{Hsinchu}
  \country{Taiwan}
}








\renewcommand{\shortauthors}{Trovato et al.}

\begin{abstract}
  Short-form video (SFV) has become a globally popular form of entertainment in recent years, appearing on major social media platforms. However, current research indicate that short video addiction can lead to numerous negative effects on both physical and psychological health, such as decreased attention span and reduced motivation to learn. Additionally, \textbf{S}hort-\textbf{f}orm \textbf{V}ideo \textbf{A}ddiction  (SFVA) has been linked to other issues such as a lack of psychological support in real life, family or academic pressure, and social anxiety. Currently, the detection of SFVA typically occurs only after users experience negative  consequences. Therefore, we aim to construct a short video addiction dataset based on social network behavior and design an early detection framework for SFVA.  Previous  mental health detection research on online social media has mostly focused on detecting depression and suicidal tendency. In this study, we propose the first early detection framework for SFVA \textit{EarlySD}. We first introduce large language models (LLMs) to address the common issues of sparsity and missing data in graph datasets. Meanwhile, we categorize social network behavior data into different modalities and design a heterogeneous social network structure as the primary basis for detecting SFVA. We conduct a series of quantitative analyses on short video addicts using our self-constructed dataset, and perform extensive experiments to validate the effectiveness of our method \textit{EarlySD},  using social data and heterogeneous social graphs in the detection of short video addiction.
\end{abstract}



\received{20 February 2007}
\received[revised]{12 March 2009}
\received[accepted]{5 June 2009}

\maketitle

\section{Introduction}
In recent years, social media platforms have introduced various short-form video (SFV) content, such as Instagram Reels, Facebook Reels, and YouTube Shorts. This is undoubtedly a commercial success inspired by TikTok, a globally popular short-form video platform. However, a potential crisis, which is named as \textbf{Short-form video addiction (SFVA)}, has emerged for users, especially adolescents.

In 2022, The Wall Street Journal introduced the concept of "TikTok Brain" \cite{TiktokBrain}, pointing out that receiving stimuli from SFV content, which does not require long-term attention, will lead to a decrease in attention span, especially for adolescents whose prefrontal cortex are not fully developed yet. In 2023, The Wall Street Journal published another article \cite{antidote} discussing whether "YouTube can be the antidote for TikTok" — since YouTube offers longer and more comprehensive video. However, the emergence of TikTok prompts other social platforms to emulate this SFV service, precluding the opportunity to solve TikTok’s negative effects by other platforms. BBC and CNN have also expressed concerns that the overuse of SFV may lead to addiction, which brings adverse effects on user’s mental health \cite{bbc,cnn}.
Current studies also indicate several negative effects of short-form videos on both physiological and psychological aspects. For instance, \cite{psy-attention} suggests that SFVA users report less interest, attention concentration, and more distractions compared to Non-SFVA users. \cite{psy-motivation} indicates that short-form video addiction has a negative impact on both intrinsic and extrinsic learning motivation. Additionally, numerous studies \cite{psy-7,psy-8,psy-9} exploring the causes of short-form video addiction have found that social phobia, school burnout, cognitive stress or lack of offline social support, and a shift towards seeking online social support can positively predict short-form video addiction, reflecting that addicted users may also be exposed to risks of psychological stress and a lack of real-life social support.

According to a survey conducted by Common Sense in 2023 \cite{commonsense-report}, which investigated the impact of social media usage habits on physical and mental health, it was found that 41\% of respondents felt that TikTok disrupted their sleep, 24\% of respondents said their daily sleep was affected, and 40\% of respondents had to stop or limit their use of TikTok because they felt it consumed too much time. Furthermore, 34\% of respondents said their lives would become worse without TikTok. Additionally, 30\% of respondents reported negative social experiences such as bullying and feeling excluded, while 31\% of respondents said that short-form video content made them feel dissatisfied with their bodies at least once a week. The above results show the inescapable negative repercussions.

The severe consequences of SFVA can be divided into three categories: \textit{physiological}, \textit{psychological}, and \textit{others}. Physiologically, SFVA may lead to issues such as sleep deprivation, resulting in a decline in work performance. Psychologically, it may include social anxiety, self-doubt, and other related problems. Other potential consequences may arise from imitating inappropriate content. Dangerous challenges have gained popularity on SFV platforms, with users becoming addicted and subsequently attempting to mimic them. For instance, the \textit{Benadryl Challenge} involves ingesting 12 to 14 antihistamine tablets to experience hallucinations, while the \textit{Blackout Challenge} encourages participants to use ropes to strangle themselves until they lose consciousness. Such challenges continue to proliferate, resulting in numerous injuries and even fatalities.

Nowadays, despite the detrimental effects of SFVA, identifying individuals afflicted with this condition early remains quite challenging for several reasons. Firstly, unlike widely recognized conditions such as depression or PTSD, the symptoms of SFVA are not well-known, and those who may be affected may not even be aware of the problem. Secondly, individuals suffering from this addiction often have a low level of insight into their condition, remaining unaware of the extent to which it is affecting their lives. Identification of SFVA may occur when vigilant supervisors, such as parents and teachers, notice concerning signs. Alternatively, it might come to light when afflicted individuals experience severe symptoms because of their addiction, prompting them to seek assistance from doctors or experts. It is only then that they come to realize that they are suffering from short-form video addiction, by which point their life has already spiraled out of control.

Based on the above points, we recognize the importance of this SFVA and its derived issue. We aim to utilize online social behavior data, which is closely associated with short-form videos and offers extensive user data, as the primary means of early detection of short-form video addiction. In this paper, we primarily focus on two directions. i) Through preliminary studies, we aim to uncover the potential relationship between online social behavior data and SFVA. We have identified differences between SFVA and Non-SFVA in certain social behavior indicators, such as searching, commenting, and bidirectional friendships. ii) Based on our preliminary results, we develop an early detection system applicable to large-scale social networks using machine learning techniques. The objective of this detection system is, given a user's social data (including behavior data and interaction patterns in social networks), to dynamically construct a social graph and extract her social attributes. We then utilize our proposed heterogeneous node classification framework to identify user labels, distinguishing specific users as SFVA or Non-SFVA.

\section*{Contributions and Impacts}
\begin{itemize}
    \item \textbf{The first SFVA real dataset with online social behavior traces.}To our best knowledge, the SFVA dataset collected in this paper is the first real SFVA dataset in the world that contains both users’ social interactions and their demographic information. This dataset bridges the gap between online social behavior data and potential risk of SFVA. We not only hope that the dataset will be beneficial for constructing early detection of short-form video addiction but also aim to raise awareness among the public and academia about the importance of SFVA.
    \item  \textbf{Quantitative analysis of SFVA users.} With the collected data of 534 samples, we conducted in-depth quantitative analyses and created portraits of users with varying degrees of addiction tendency to short-form videos, including \textit{SFVA}, \textit{Potential-SFVA}, and \textit{Non-SFVA}. This is achieved by collecting personal social media data (Instagram). We uncovered the correlation between addictive behaviors and social media usage patterns. In fact, our machine learning framework based on a deep understanding of social behavior data, is greatly benefited from the real user data, enabling us to predict and understand the addiction more accurately and comprehensively.
    \item \textbf{Novel framework for SFVA early detecton.} We developed a novel framework, named \textit{Early-state Short-Form Video Addiction Detector (EarlySD)}, for the early SFVA detection. The framework addresses challenges such as data sparsity and feature incompleteness. We leveraged social behavior patterns and personal information of users, employing large language Models (LLMs) to overcome issues related to data sparsity. Our design includes a heterogeneous graph neural network classifier that fully considers the characteristics of users and their interests in specific topics to determine their propensity for SFVA. The extensive experiments using real-world Instagram data verify that \textit{EarlySD} outperforms state-of- the-art methods in terms of all the effectiveness measures.
\end{itemize}

\section{Related Works}
\subsection{Mental Health Detection with Social Media Data}

Nowadays, social media platforms like Facebook (now Meta), Instagram, YouTube, and Twitter provide an online environment for information sharing, emotion expression, and communication with various types of interaction. Consequently, multiple forms of user-generated data, including text, photos, videos, and social interaction behavioral data, become potential to represent specific individuals and are utilized in various applications. Mental health detection is one of the application tasks that benefit from social media data. In recent years, massive data of user contents, user activities, and interpersonal networks on social media have been harvested to serve as identifiable features in several frameworks for depression and PTSD detection. Specifically, deep neural networks (DNNs) show notable success in previous detection works \cite{DepressionNet, MentalNet, MentalSpot, SNMDD, ReadAndSee, AudiBERT, COMMA}.

Most recent studies have utilized content from social media, such as text or photos. For instance, Mann et al. proposed multimodal learning from users’ posts, including feature extraction from captions and images. The feature representations from two different modalities are concatenated to train a detector for depression \cite{ReadAndSee}. MentalSpot \cite{MentalSpot} used a triplet network to extract users’ embeddings based on their written content. For each target user, the derived embeddings from his/her top-k homogenous friends are fully utilized to train a downstream depression detection model. DepressionNet \cite{DepressionNet} selects depression-relevant content initially based on BERT-BART, which is integrated with a convolution neural network (CNN) and attention-based GRU to perform depression prediction. MentalNet \cite{MentalNet} utilized various interaction types on social media and constructed a heterogeneous graph for each user. The graph, representing different interaction modes, was treated as input in a graph classification model to detect users’ depression states. Other works like SNMDD \cite{SNMDD} extract social interaction features and detect social network mental disorders such as cyber-relationship addiction, net compulsion, and information overload using multi-source semi-supervised learning.

In summary, we can observe the extensive application of social media data in the field of mental health. The content generated by users and their behavior patterns on social media are crucial for training an effective detector for mental health issues. However, the above-mentioned studies have certain limitations. Most mental health detectors trained with social media data primarily focus on depression. Considering that different mental disorders emphasize different social features and content, these detection frameworks cannot be directly applied to our proposed SFVA early detection problem. To the best of our knowledge, we are the first to construct an SFVA dataset and propose a study that simultaneously leverages social features and user interest topics to build a heterogeneous graph-based classifier for SFVA detection.

\subsection{Graph Neural Network on Social Network}
Graph Neural Networks (GNNs) have emerged as a powerful framework for learning representations of nodes in graph-structured data, as GNNs can naturally integrate node information and topological structure. Online social network data is well-suited for solving problems using GNNs for several reasons:\\ 

\noindent \textbf{Intrinsic graph structure.} Social networks are naturally represented as graphs, where nodes represent users and edges represent relationships or interactions between users. 

\noindent \textbf{Relational information.} In social networks, the connections between users carry significant information. GNNs can effectively aggregate and propagate information through these connections, capturing the complex dependencies and influence patterns within the network.\\

By leveraging the unique strengths of GNNs, a wide range of problems in social networks can be solved, from user recommendation systems to detecting anomalous behaviors, and beyond. The fundamental idea behind GNNs is to iteratively update node representations by aggregating information from neighboring nodes. One of the pioneering works in this field is the Graph Convolutional Network (GCN) proposed by Kipf and Welling \cite{GCN}, which introduces a convolutional operation on graphs by propagating information from neighboring nodes using the graph Laplacian. Another influential model is the Graph Attention Network (GAT) \cite{GAT}, which utilizes attention mechanisms to learn adaptive weights for aggregating node features, allowing it to selectively attend to important neighbors during message passing. Beyond GCN and GAT, several other classic extensions of GNNs have been proposed. GraphSAGE (Graph Sample and Aggregation) \cite{GraphSage} performs inductive learning by sampling and aggregating features from a node's local neighborhood, enabling scalable representation learning on large graphs. The Graph Isomorphism Network (GIN) \cite{GIN} generalizes graph neural networks to be as powerful as the Weisfeiler-Lehman graph isomorphism test, achieving state-of-the-art performance on various graph tasks. HetGNN \cite{HetGNN} is proposed for heterogeneous graphs that are composed of various types of nodes and edges.

\subsection{Large Language Model as Enhancer on Graph-Related Task}
When training node representations with text features in Graph Neural Networks (GNNs), traditional Natural Language Processing (NLP) methods such as bags-of-words, TF-IDF, and skip-grams are widely adopted. However, with the rise of \textbf{Large Language Models (LLMs)} in recent years, there has been growing interest in applying LLMs on graphs, particularly for tasks related to Graph Neural Networks. Indeed, the powerful reasoning capabilities and understanding of text exhibited by LLMs can serve as enhancers to assist in GNN training \cite{LLMforGraph1, LLMforGraph2}. Specifically, when dealing with nodes in our target graph that possess text-type features, LLMs, with their understanding of text features, can capture deeper information about these nodes compared to GNNs, which excel in capturing structural information. When combined, the information captured by LLMs improves the effectiveness of GNNs, resulting in better-quality node representations for downstream tasks. For example, \textbf{TAPE} \cite{TAPE} proposes to use LLMs to enhance the node features of text-attribute graphs (TAGs) to improve performance in downstream tasks. \textbf{KEA} (Knowledge-Enhanced Augmentation) \cite{KEA} introduces two strategies for LLMs to enhance text attributes: at the feature level and text level, optimizing node embedding for both embeddings and original text information. Furthermore, \textbf{LLMRec} \cite{LLMREC} extends the possibilities of LLMs. In addition to leveraging node text features to enhance the understanding of item and user node attributes, it strengthens the connections between users, augmenting sparse graphs for better performance in downstream recommendation tasks.

These works demonstrate that LLMs not only improve the quality of textual data on Graph Neural Networks but also tackle potential issues on real-world graphs, such as sparsity and missing features. In our work, we also aim to predict SFVA via social network graphs, which exhibit these sparse and text-rich node characteristics. We believe that incorporating LLM assistance can provide us with a more comprehensive graph for feature training, leading to better performance in downstream tasks.

\section{Preliminary Analysis}
While previous research on mental health detection has demonstrated the significant effectiveness of diverse social media data in addressing issues such as depression and anxiety, as we have mentioned, there is still no study validating the use of social media behavior data as an effective approach for the problem of short video addiction. Therefore, we hope to first understand the relationship between personal features, social features, and SFVA before constructing the SFVA early detector. We aim to clarify the potential to identify the tendency of SFVA through social media data.

\subsection{Real Data Collection}
We recruited 534 Taiwanese participants aged 18 and above, asking them to complete a comprehensive questionnaire. The questionnaire included the Short Video Addiction Scale, derived from Professor Sue-Huei Chen’s Chinese Internet Addiction Scale (CIAS) \cite{CIAS}, the Social Interaction Anxiety Scale \cite{SIAS}, and the Big Five Personality Scale \cite{Big5}. In the pilot study, volunteer participants were invited to provide their social media footprints (Instagram). Our aim was to explore potential addiction-related characteristics through social media footprints and to observe the differences in video viewing behaviors between SFVA and Non-SFVA through these digital footprints.
From February 2, 2024, to April 15, 2024, we received a total of 534 responses. We excluded 32 responses that did not pass the questionnaire's lie detection items. Our final sample consists of 357 females and 111 males with a median age of 22, with 275 participants currently in school.

\subsection{Feature Extraction}
\noindent\textbf{Personal Demographic Feature.} Personal features include demographic data such as age, gender, educational stage, the number of short video platforms used, as well as quantified indicators of personality traits obtained through scales, such as the Big Five Personality traits and social anxiety.

\noindent\textbf{Social Behavior Feature.} We categorize the voluntarily provided Instagram data of participants into four major types. (1) \textbf{Connection}: The users’ followers and followings, which generally indicate friendship on the social media platform. (2) \textbf{Textual content}: User can share content in various form on social platforms, including text, image and reels. In particular, the Reels service on Instagram is an example of short-form video (SFV) content. (3) \textbf{Search and Topic}: The preferences exhibited by users on social platforms, such as content searching or content recommended by the platform itself.(4) \textbf{Interaction}: Social media platforms typically offer various ways to interact. On Instagram, users can interact with others through likes, comments, and private messages. 

\subsection{Quantitative Insight from demographic and social behavior feature}
Through our preliminary study, we found that both personal features and social features possess potential efficacy in capturing short video addiction. We utilized ANOVA test to analyze all features and identify features demonstrating a significant difference between SFVA and Non-SFVA groups.

\begin{table}[ht]
\centering
\caption{Summary of Statistically Significant Features}
\label{tab:statistical_findings}
\resizebox{\linewidth}{!}{%
\begin{tabular}{@{}llp{7cm}l@{}}
\toprule
Type & Variable & Description & p-value \\
\midrule
Personal & Age & Younger users show higher SFVA tendency. & 0.02* \\
 & Social Anxiety & Higher social anxiety is associated with higher SFVA tendency. & 0.001** \\
 & Neuroticism & Higher neuroticism is associated with higher SFVA tendency. Neuroticism reversely reflects the emotional stability. & 0.002** \\
\addlinespace 
\midrule
 Social & Searching & SFVA users show more searching behaviors on social media platforms. & 0.03* \\
 & Leaving Comment & SFVA users show more commenting behaviors on social media platforms. & 0.05 \\
 & Bi-directional Friendship & SFVA users tend to have a higher proportion of bi-directional friendships, defined as the ratio of mutual follows to total follows on social media platforms. & 0.04* \\
\bottomrule
\end{tabular}%
}
\begin{center}
\footnotesize
* means p-value < 0.05, ** means p-value < 0.01.
\end{center}
\end{table}

\begin{figure}[H]
    \centering
    \includegraphics[width=\linewidth]{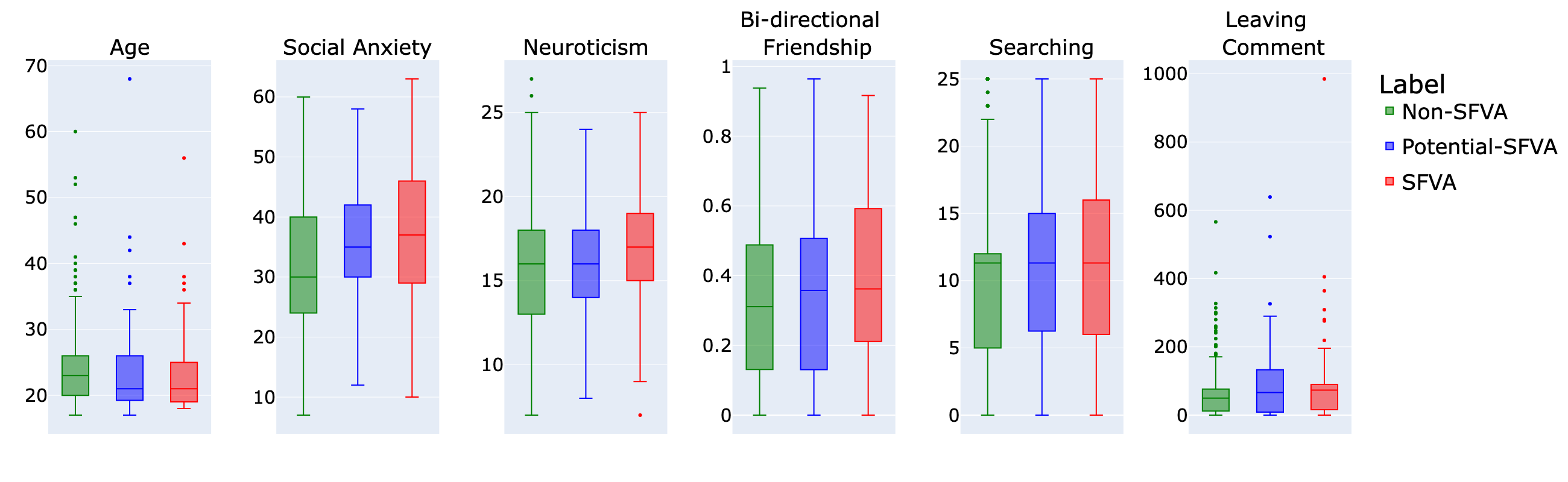}
    \caption{Significant difference between SFVA and Non-SFVA shows age, social anxiety tendency and personality are potentially effective features.}
    \label{fig:Prelimianry-FeatP}
\end{figure}

\noindent\textbf{Insights from personal features.} Table\ref{tab:statistical_findings} and figure \ref{fig:Prelimianry-FeatP} shows that personal feature and social feature are both potentially important to distinguish SFVA and Non-SFVA.  
(1) There was a significant difference in age among SFVA and Non-SFVA (p = 0.02), indicating that compared to Non-SFVA, SFVA show a trend towards a younger age.
(2) There was a significant difference in social interaction anxiety among SFVA and Non-SFVA (p = 0.001). Additionally, there was a significant difference in Neuroticism in the Big-5 personality traits between SFVA and Non-SFVA (p = 0.002). This suggests a significant association between the tendency for SFVA and the psychological state of social media users. In fact, numerous previous psychological studies \cite{PsyPersonality1, PsyPersonality2, PsyPersonality3, PsyPersonality4} have indicated the relationship between personality traits, social anxiety, and addiction issues.

\noindent\textbf{Insights from social behavior features} Table\ref{tab:statistical_findings} and figure \ref{fig:Prelimianry-FeatP} also shows the fact that:
(1) On social media platforms, \textbf{SFVA have a habit of searching for other accounts (or called social searching) especially in searching for their own friends, which significantly differs from Non-SFVA}: Searching on social platforms or search engines demonstrates specific actions driven by interest in certain people or things. We found that SFVA exhibit a more pronounced interest in others (especially their friends) on social media platforms (p = 0.03).
(2) \textbf{SFVA are more active in various type of interaction on social platform.} Social platforms offer various interaction mechanisms. Taking Instagram as an example, liking, commenting, and stories are opportunities for user interaction. We found that short video addicts are more active in commenting and stories interactions, with a nearly significant difference in interaction frequency compared to Non-SFVA (p = 0.05).
(3) Observing the social status on social media platforms, SFVA and Non-SFVA showed no difference in the number of social connections, but \textbf{SFVA users tend to establish bidirectional friendship edges}: On platforms like Instagram, the friend mechanism is divided into following and followers. Typically, mutual following is considered a closer connection between both parties, rather than a one-sided following behavior. We found that SFVA users are more inclined to establish such bidirectional connections (p = 0.04).

\noindent\textbf{Insights from users and their interested content.}
Another interesting finding is the relationship between the types of topics users are interested in and their addiction tendencies. In addition to users' personal characteristics, which can affect their tendency toward addiction, the content recommended by the platform is also a key factor influencing users' dwell time. In analyzing the data provided by Instagram, we found a dataset of customized recommended content based on users' browsing behavior and records on social media. By observing the distribution of SFVA and Non-SFVA users across these topics, we found that the proportion of SFVA users varies across different topics. This suggests that the content users encounter on social media or short video platforms also has the potential to influence their addiction tendencies.

We use Jensen–Shannon divergence (JSD) to calculate interested topic divergence. JSD provides a robust metric for quantifying the similarity between two probability distributions, making it particularly useful for analyzing interest topics based on their frequency distributions between SFVA, Potential-SFVA, and Non-SFVA. Specifically, we take sets of interest topics from SFVA and Non-SFVA users, compute their respective frequency distributions, \(P\) and \(Q\). The JSD is then calculated based on the average distribution \(M\), which is the mean of \(P\) and \(Q\). Mathematically, JSD is computed as:

\begin{equation}
    \text{JSD}(P \| Q) = \frac{1}{2} D_{KL}(P \| M) + \frac{1}{2} D_{KL}(Q \| M) \in [0, 1]
\end{equation}

where \(D_{KL}\) denotes the Kullback-Leibler divergence. This measure captures both the divergence of \(P\) and \(Q\) from their average distribution \(M\). By evaluating the JSD, we can determine how similar or different the interest topic distributions are, which is crucial for understanding patterns and preferences among users. A lower JSD value indicates that the interest topic distributions are more similar, while a higher value reflects greater divergence. Table \ref{tab:jsd_divergence} shows the divergence between SFVA, Potential-SFVA, and Non-SFVA. We can observe that SFVA and Non-SFVA show larger divergence while Potential-SFVA and SFVA show smaller divergence.

\begin{table}[h!]
    \centering
    \caption{Jensen–Shannon Divergence between Non-SFVA, Potential-SFVA and SFVA}
    \label{tab:jsd_divergence}
    \begin{tabular}{|l|c|}
        \hline
        Group & JSD \\
        \hline
        Non-SFVA \& SFVA & 0.48 \\
        Non-SFVA \& Potential-SFVA & 0.41 \\
        SFVA \& Potential-SFVA & 0.39 \\
        \hline
    \end{tabular}
\end{table}

Through this preliminary study, we observed that certain personal and behavioral features on social media are useful in identifying SFVA, Potential-SFVA, and Non-SFVA users. SFVA users are particularly active in specific social behaviors, such as social searching and commenting. Additionally, they exhibit a significant inclination towards bidirectional friendships in their social interactions. These findings regarding the association of social indicators with SFVA allow us to preliminarily validate the potential effectiveness of using social network data—the most common platform for short-form video—in the early detection of SFVA.

\section{Problem Formulation}
Our goal is to utilize social network data to detect individuals suffering from Short-Form Video Addiction (SFVA). We frame this machine learning problem as a node classification task with graph refinement. We denote the directed input social graph as $G = (V, E, X)$, where $V$ represents the node set (individuals) of size $N$, and $E \subset V \times V$ represents the edge set (social connections between individuals). We use $X \in \mathbb{R}^{N \times F}$ to represent input node features with $F$ dimensions, which encompass the textual, social, and SFVA-related attributes associated with each individual. We further denote the adjacency matrix as $A \in \{0, 1\}^{N \times N}$, where $a_{ij} = 1$ if there exists an edge between nodes $v_i$ and $v_j$. In the social graph, $N(i)$ denotes the set of all 1-hop neighbors of node $i$.

The graph refinement task is defined as follows: Given a graph $G = (V, E, X)$, which contains an incomplete set of edges $E$ and incomplete attributes $X$, we aim to find a predictive function $F_{\text{GSL}}$ that takes $G$ as input and outputs its predicted edge set $E'$ and complete attributes $X'$, such that the difference between the new graph $G' = (V, E', X')$ and the ideal graph $G_{\text{IDL}} = (V, E_{\text{IDL}}, X_{\text{IDL}})$ is minimized. Here, the ideal graph $G_{\text{IDL}} = (V, E_{\text{IDL}}, X_{\text{IDL}})$ has all the edge and attribute information.

For the downstream node classification task, we take the refined graph $G' = (V, E', X')$ as input. Our objective is to develop a classifier function $F_{\text{C}}$ to determine whether a node label is SFVA or Non-SFVA.

\section{METHODOLOGY}
\subsection{EarlySD Framework}
\begin{figure*}
  \includegraphics[width=\textwidth]{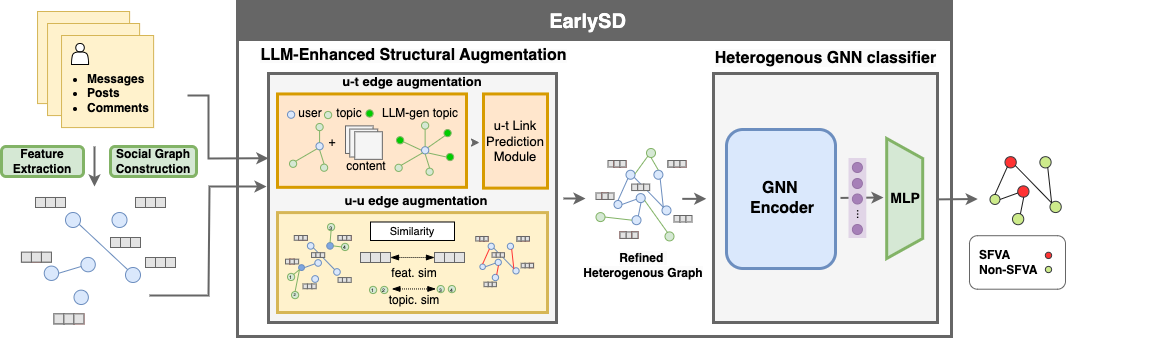}
  \caption{\textit{EarlySD} is comprised of two main components: (1) social network graph construction and LLM-enhanced structural augmentation and (2) heterogeneous graph neural network encode and an MLP projection head.}
  \label{fig:earlySD}
\end{figure*}

We proposed \textit{EarlySD} as our SFVA detector in figure \ref{fig:earlySD}, which utilized heterogeneous Graph Neural Network (GNN) as an encoder to learn user embeddings. The well-trained embedding is then adopted in a classifier to distinguish whether users exhibit SFVA tendencies. In this heterogeneous framework, considering the preliminary study's findings that SFVA users and Non-SFVA users show different content preferences on social platforms, we introduced topics as an additional node type. This inclusion aims to capture users' interests during the embedding update process.
As typical real-world datasets, our collected SFVA dataset also suffers from severe sparsity and incomplete feature issues. Moreover, we face significant challenges due to the small sample size. To address these problems, we introduced a large language model (LLM) as an enhancer to optimize our dataset with edge augmentation. Then we utilize the refined social graph into heterogeous GNN classifier to determine SFVA tendancy for each node.

\subsection{LLM-enhanced Structural Augmentation}
In our constructed social graph, there are two types of nodes: users and topics of interest to these users. The graph includes two types of edges:

\begin{itemize}
    \item \textbf{User-to-User Edge (u-u edge)}: Typically, in a social graph, these edges represent relationships between users, such as friendships, follow relationships, or interaction records on social networks.
    \item \textbf{User-to-Topic Edge (u-t edge)}: In our heterogeneous structure design, the user-to-topic edges represent a user's interest in specific topics.
\end{itemize}

\subsubsection{u-u Edge Augmentation}
Graph Neural Networks (GNN) propagate node information to a specific target node from neighboring nodes. The adjacency matrix on a graph is often predefined, such as friendship in social networks or citation relationships in citation networks. However, there is no predefined edge between users in our dataset. As a result, we augment the topological structure with user similarity. Here, we consider both user feature similarity and their interested topic similarity. Feature similarity $sim_f$ can be intuitively defined as follows,

\begin{equation}
    sim_f = sim(f_u, f_v)
\end{equation}

where $f_u$ and $f_v$ represent the features of users $u$ and $v$, respectively, and $cos(\cdot)$ denotes cosine similarity. For topic similarity, we introduce an LLM module with user interest topic lists, which are originally derived from the Instagram recommendation system. We define topic similarity as,

\begin{equation}
    sim_t = LLM(T_u, T_v) \in [-1, 1]
\end{equation}

where $T_u$ and $T_v$ represent the topic lists for users $u$ and $v$, respectively. The topic similarity $sim_t$ ranges from -1 to 1 to determine whether the topics of interest between the two users are similar.

The overall similarity score is defined as $\alpha \cdot sim_f + (1 - \alpha) \cdot sim_t$. We introduce a Kolmogorov-Arnold Network (KAN) to learn the weight parameter $\alpha$, which controls the importance of user feature similarity and interested topic similarity.

\subsubsection{u-t Edge Augmentation}
To address the issue of incomplete user-topic edges in the input social graph, we adopt a two-step approach:
\begin{enumerate}
    \item \textbf{Topic Set Extension with LLM}: Recognizing that the initial topic set may not encompass all user interests, we incorporate a Large Language Model (LLM) module to generate additional topics. The LLM analyzes users' content (such as comments and stories) to identify potential interests, thereby expanding the original topic set.
    
    \item \textbf{GNN Link Prediction for u-t Edge}: Users who have already provided u-t edge information serve as training data for the heterogeneous link prediction module between user nodes and topic nodes. We train the link prediction model using the following procedure:
    
    \begin{equation}
        \mathcal{E}_{ut} = \sigma(\cos(h_u, h_t))
    \end{equation}
    
    \begin{equation}
        h_u^l = HGC_{lp}(h_u^{l-1}), \quad h_t^l = HGC_{lp}(h_t^{l-1})
    \end{equation}

    \begin{equation}
    \mathcal{L}_{lp} = BCE(\mathcal{E}_{ut}, y_{ut})
    \end{equation}
    
    Here, \(\mathcal{E}_{ut}\) represents the probability of an edge existing between user \(u\) and topic \(t\), calculated using the cosine similarity between their embeddings, followed by a sigmoid function \(\sigma\). \(HGC_{lp}(\cdot)\) denotes a heterogeneous GNN layer, and \(h_u\) and \(h_t\) are the embeddings of user \(u\) and topic \(t\), respectively. \(BCE(\cdot)\) is the binary cross-entropy loss function used to update the link prediction module.
\end{enumerate}

This augmentation process not only generates potential topics but also constructs new u-t edges, thereby enriching the user-topic relationship in the social graph.

\subsection{Heterogenous Graph-based SFVA classifier}
\subsubsection{Embedding Initialization}

In our preliminary analysis, we have identified that both personal features and social features significantly contribute to capturing individual tendencies towards SFVA. Therefore, we concatenate these features and define the initial embedding as follows:

\begin{equation}
    h_u^0 = FFN_u(X_u^p || X_u^s)
\end{equation}

where \( || \) represents the concatenation operation, \( X_u^p \) denotes the personal features of user \( u \), and \( X_u^s \) denotes the social features of user \( u \). \( FFN_u \) is a feed-forward neural network to obtain initial embedding for user node.

To preserve the original semantics of the topic nodes, we utilize the dictionary in the large language model (LLM) to obtain text embeddings \( X_t \in \mathbb{R}^{d_t} \) for each topic, where \( d_t \) represents the dimensionality of the LLM dictionary. We subsequently apply a transformation, \( FFN_t \), to these topic embeddings to obtain transformed topic embeddings \( h_t^0 \), ensuring that the dimension of \( h_t^0 \) aligns with the dimension of the user node embeddings.

\begin{equation}
    h_t^0 = FFN_t(X_t)
\end{equation}

\subsubsection{Heterogeneous Graph Neural Network Classifier as SFVA Early Detector}
We introduce a GNN designed to encode the heterogeneous social network graph for learning node embeddings. The GNN encoder incorporates two message propagation functions to represent information flow between user-user and user-topic interactions. User embeddings are learned through propagation as follows:

\begin{equation}
h_u^{(l)} = \sum_{v \in N_u^{\text{user}}} f_{uu}(h_v^{(l-1)}) + \sum_{t \in N_u^{\text{topic}}} f_{ut}(h_t^{(l-1)})
\end{equation}

Here, \( h_u^{(l)} \) denotes the user embedding in the \( l \)-th layer, \( f_{uu} \) represents the message propagation function between user nodes, and \( f_{ut} \) represents the message propagation function between users and their interested topics. \( v \) denotes a user neighbor, while \( t \) denotes a topic neighbor of \( u \).

The node embeddings generated by the encoder are subsequently utilized by an MLP classifier to predict tendencies towards SFVA.

\section{Experiment}
In this section, we empirically demonstrate the effectiveness and superiority of the proposed \textit{EarlySD} for SFVA detection using our collected real SFVA dataset. As mentioned in our preliminary study, we collected the first SFVA dataset based on social media. The relevant statistics are detailed in the table. First, we briefly describe the implementation details, and then compare the SFVA detection problem with baselines, including classic classifiers and three state-of-the-art mental health detection frameworks. Finally, we conduct extensive ablation studies to verify the effectiveness of our refinement module and main classification module.

\subsection{Data Statistics}
We recruited 534 participants to respond to the SFVA scale, derived from Professor Sue-Huei Chen's Chinese Internet Addiction Scale (CIAS) \cite{CIAS}. Scores below 58 are classified as Non-SFVA, scores above 64 indicate SFVA symptoms, and scores ranging from 58 to 64 are categorized as Potential-SFVA, suggesting a predisposition towards addiction though not sufficiently severe to be classified as SFVA. In our study, we amalgamated SFVA and Potential-SFVA cases, treating them as positive samples potentially at risk of SFVA's adverse effects. We excluded 32 invalid entries due to severe data incompleteness or failure to pass the catch questions, thereby mitigating data noise. Table \ref{tab:geneal-stats} and \ref{tab:content-stat} presents statistics for the remaining 468 valid participants, including their average SFVA scores, the number of topics they are interested in, and their textual social media contents.
\begin{table}[ht]
\centering
\caption{Summary statistics our collected SFVA dataset}
\label{tab:geneal-stats}
\begin{tabular}{@{}lcccc@{}}
\toprule
& \multicolumn{1}{c}{\textbf{Number of User}} & \multicolumn{3}{c}{\textbf{SFVA Score}} \\
\cmidrule(lr){2-2} \cmidrule(lr){3-5}
\textbf{Group} & \textbf{Sum} & \textbf{Min} & \textbf{Max} & \textbf{Mean} \\
\midrule
Non-SFVA  & 259 & 26 & 57 & 44.73 \\
SFVA  & 134 & 64 & 95 & 70.8 \\
Potential-SFVA  & 75 & 58 & 63 & 60.11 \\
\bottomrule
\end{tabular}
\end{table}

\begin{table}[ht]
\centering
\caption{Summary statistics for textual-related behavior on social media}
\label{tab:content-stat}
\resizebox{\linewidth}{!}{%
\begin{tabular}{@{}lccccccccc@{}}
\toprule
& \multicolumn{3}{c}{\textbf{Posts}} & \multicolumn{3}{c}{\textbf{Stories}} & \multicolumn{3}{c}{\textbf{Comments}} \\
\cmidrule(lr){2-4} \cmidrule(lr){5-7} \cmidrule(lr){8-10}
\textbf{Group} & \textbf{Min} & \textbf{Max} & \textbf{Mean} & \textbf{Min} & \textbf{Max} & \textbf{Mean} & \textbf{Min} & \textbf{Max} & \textbf{Mean} \\
\midrule
Non-SFVA  & 2 & 1214 & 49.87 & 2 & 3876 & 474.08 & 1 & 6727 & 541.75 \\
SFVA  & 2 & 416 & 27.41 & 2 & 4603 & 640.53 & 1 & 27137 & 697.39 \\
Potential-SFVA  & 2 & 507 & 47.35 & 15 & 2812 & 621.78 &  8 & 2277 & 481.19\\
\bottomrule
\end{tabular}%
}
\end{table}

\subsection{Baselines}
In order to demonstrate the effectiveness and superiority of our approach, we compare our framework \textit{EarlySD} with six methods.

\subsubsection{Feature-based classifier}
\begin{itemize}
    \item \textbf{Traditional machine learning methods:} We compare \textbf{AdaBoost, Random Forest, KNeighbors, and SVC RBF kernel as classifiers} with extracted user features. We took the method with the best performance among all and denoted it as \textbf{ML-Best} as one of the baselines.
    \item \textbf{Multi-layer Perceptron:} We utilized a 2-layer MLP with extracted user features as input.
\end{itemize}

\subsubsection{State-of-the-art methods for mental health detection}

Since SFVA detection is still a novel mental health issue, there is no customized framework designed for it yet. We chose three similar frameworks that also utilized social media data in various ways and compared the performance with our \textit{EarlySD} using the same self-collected SFVA dataset.

\begin{itemize}
    \item \textbf{ReadAndSee} (ICWSM’20) \cite{ReadAndSee} proposed a multimodal learning framework using users’ posts, including feature extraction from post captions and images. The feature representations from two different modalities are concatenated to train a detector for depression. In our setting, we used social features to replace image features. The provided source code utilized pre-trained ELMo in Portuguese to generate text embedding. We utilized pre-trained BERT to replace ELMo since our user content is all in Chinese.
    \item \textbf{MentalSpot} (CIKM’21) \cite{MentalSpot} employed contrastive learning and implemented dynamic mean shift pruning to identify the most homogeneous friends. The text content from each user and her most homogeneous friends are used for depression detection. We retrieved the top-3 most homogeneous friends for each user in our setting.
    \item \textbf{DepressionNet} (SIGIR’21) \cite{DepressionNet} proposed a depression detection framework by selecting relevant content through a hybrid extractive and abstractive summarization strategy. The fine-grained and relevant content then goes into a deep learning framework comprising a unified learning machinery comprising Convolutional Neural Networks (CNN) coupled with attention-enhanced Gated Recurrent Units (GRU) models.
\end{itemize}

\subsection{Evaluation Metrics}
To evaluate the model performance, we adopt accuracy (ACC), precision (PRE), recall (REC), and F1-score as metrics. The definitions of these metrics are as follows: 
\begin{enumerate}
    \item \text{ACC} = $\frac{\text{TP} + \text{TN}}{\text{TP} + \text{TN} + \text{FP} + \text{FN}}$
    \item \text{REC} = $\frac{\text{TP}}{\text{TP} + \text{FN}}$
    \item \text{PRE} = $\frac{\text{TP}}{\text{TP} + \text{FP}}$
    \item \text{F1-score} = $\frac{2 \times \text{PRE} \times \text{REC}}{\text{PRE} + \text{REC}}$
\end{enumerate}
Here, TP, TN, FP, and FN denote true positive, true negative, false positive, and false negative, respectively.

\subsection{Performance Comparison}
Among all methods, our heterogenous framework, \textit{EarlySD}, demonstrates the best performance in detecting short-form video addiction (SFVA). Compared to traditional machine learning classifiers and Multilayer Perceptron, which are feature-based, \textit{EarlySD} fully considers the information between 2-hop neighbors. In comparison with general graph-based methods (GCN, GAT), the heterogenous framework not only encodes the information between users but also captures the influence of topics on users.

When compared with three state-of-the-art depression detection frameworks, we observe that their performance is similar to general feature-based methods. This is likely because these methods primarily use the user's content as the main classification feature. For instance, \textbf{MentalSpot}, which only considers the textual content of the user and their top-k homogeneous friends, performs almost at random. Although \textbf{ReadAndSee} and \textbf{DepressionNet} consider some social features in addition to textual content, they do not fully utilize the quantitative features derived from various interaction types on social media. They focus solely on posting-related behavior features.

This observation indicates that different mental health problem detection tasks emphasize different features. Maximizing the utilization of these features also varies: for user content, textual content and posting style are beneficial for depression detection, while in SFVA detection, extracting the user's topics of interest from textual content is more important. Additionally, for social features, it is more effective to explore behaviors across different interaction types and convert them into quantitative indicators, rather than focusing solely on posting behavior indicators, to effectively detect SFVA tendency.

\begin{table}[ht]
\centering
\caption{Overall performance comparison on SFVA detection}
\label{table:results-binary}
\resizebox{\linewidth}{!}{%
\begin{tabular}{lcccc}
\toprule
\textbf{Method} & \textbf{ACC (\%)} & \textbf{PRE (\%)}& \textbf{REC (\%)}& \textbf{F1 (\%)} \\ 
\midrule
\textbf{ML-Best}   & 62.50 & 62.50  & 62.36 & \underline{62.32} \\ 
\textbf{MLP}        & 58.33 & 58.52 & 57.99 & 57.51 \\ 
\textbf{GCN}        & \underline{65.71} & \underline{63.64} & 67.23 & 62.05 \\ 
\textbf{GAT}        & 60.00 & 58.99 & \underline{68.33} & 53.60  \\ 
\textbf{ReadAndSee (ICWSM'20)} & 58.46 & 56.43 & 56.71 & 56.44 \\ 
\textbf{MentalSpot (CIKM'21)} & 50.77 & 47.17 & 46.48 & 46.17 \\ 
\textbf{DepressionNet (SIGIR'21)} & 57.58 & 51.72 & 51.72 & 51.72 \\ 
\midrule
\textbf{EarlySD (ours)} & \textbf{80.00} & \textbf{74.29}& \textbf{83.87}& \textbf{78.79}\\ 
\bottomrule
\end{tabular}%
}
\end{table}

\subsection{Ablation Study}

\subsubsection{Expanded Topic Set}
In the social graph refinement process, an important step is expanding the potential topic set based on user-generated content. We had 317 topics in the original set, and we expanded the topic set with an additional 51 topics with the help of Large Language Module. From table \ref{tab:wo-expandedtopic}, we could observe better performance with the extended topic set, indicating that user content has a significant potential to reflect their interests.

\begin{table}[ht]
\centering
\caption{Performance comparison w/o expanded topic set}
\label{tab:wo-expandedtopic}
\resizebox{\linewidth}{!}{%
\begin{tabular}{@{}lcccc@{}}
\toprule
\textbf{Topic Set Selection} & \textbf{ACC(\%)} & \textbf{PRE(\%)} & \textbf{REC(\%)} & \textbf{F1(\%)} \\ \midrule
\textbf{Expanded (368)}  & \textbf{80.00} & \textbf{74.29}& \textbf{83.87}& \textbf{78.79}\\
\textbf{Original (317)} & 78.57 & 71.43& 83.33& 76.92\\
\bottomrule
\end{tabular}%
}
\end{table}

\begin{figure}[h]
\centering
\subfigure[Original topic set]{
\label{Fig.sub.1}
\includegraphics[width=0.48\linewidth,height=3cm]{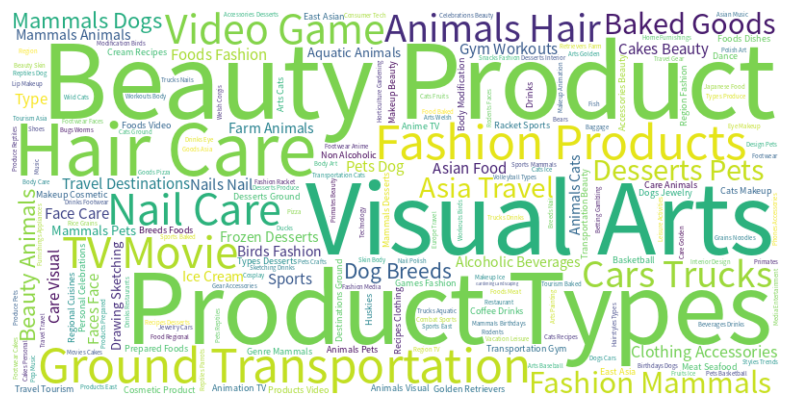}}
\subfigure[Expanded topic set]{
\label{Fig.sub.2}
\includegraphics[width=0.48\linewidth,height=3cm]{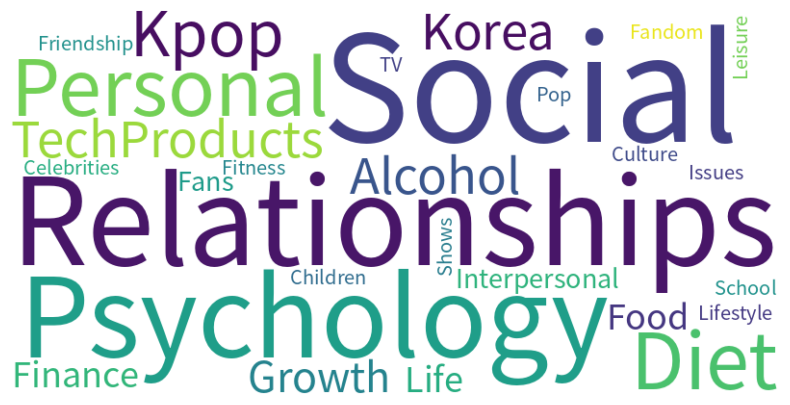}}
\caption{Expanded topic set capture additional information from user generated content}
\label{1}
\end{figure}

Figure \ref{Fig.sub.1} shows the example of the original topic set, while Figure \ref{Fig.sub.2} shows the additional topics derived from user-generated content with the help of the LLM module. We can observe that the new topic set covers more abstract topics, including relationships, friendships, and personal growth. For instance, K-pop, short for Korean Pop, as a specific music genre originating from South Korea, is also detected by the LLM and added to our topic set.

\subsubsection{Structural Augmentation with u-u Edge and u-t Edge}
In structural augmentation process, we attempted to supplement potential edges in a heterogeneous social graph with severe sparsity. We have two edge types: u-u edges are edges between users, while u-t edges denote edges between users and topic nodes. Before augmentation, our heterogeneous social graph contained only 6,621 originally existing u-t edges. We show the refinement effectivensess in table \ref{tab:edgeRefinement} With u-u edge refinement, where we added edges between users based on similarity, we improved test accuracy from 68.57\% to 71.43\%, demonstrating the effectiveness of similarity-based edge addition for enhancing connectivity. On the other hand, we improved test accuracy to 77.14\% with u-t edge refinement, which underscores the importance of handling missing information. By applying both edge refinement techniques, we achieved the best performance on the SFVA detection problem.

\begin{table}[ht]
\centering
\caption{Effectiveness of structural augmentation in EarlySD.}
\label{tab:edgeRefinement}
\resizebox{\linewidth}{!}{%
\begin{tabular}{@{}lcccc}
\toprule
\textbf{Edge Refinement}& \textbf{ACC(\%)} & \textbf{PRE(\%)} & \textbf{REC(\%)} & \textbf{F1(\%)} \\ \midrule
\textbf{w/o u-u \& u-t}& 68.57& 54.29& 76.00& 63.33\\
\textbf{w/o u-u}& 71.43& 60.00& 77.78& 67.74\\
\textbf{w/o u-t}& 77.14& 68.57& 82.76& 75.00\\
\midrule
\textbf{All}& \textbf{80.00}& \textbf{74.29}& \textbf{83.87}& \textbf{78.79}\\
\bottomrule
\end{tabular}%
}
\end{table}

\subsubsection{Different Feature Modality}
As we mentioned in preliminary study, we divided user feature into five different modalities as P (Personal), C (Connection), T (Textual content), S (Search) and I (Interaction). P belongs to personal demographic feature, while the other four belong to social behavior feature. The details of the features could be found in Appendix. Here we use this ablation study to show all these modalities are important for SFVA detection. With single modality (P, C, T, S or I) alone, the performance is worse than utilizing multiple modalities. We could also observe that interaction (I) feature is more effective than other social behavior features.  The best performance appeared when we fully utilize both personal demographic features and social behavior features.
\begin{table}[ht]
\centering
\caption{Performance metrics with different feature modality}
\label{tab:similaritysetting}
\resizebox{\linewidth}{!}{%
\begin{tabular}{@{}lcccc@{}}
\toprule
\textbf{Modalities}& \textbf{ACC(\%)} & \textbf{PRE(\%)} & \textbf{REC(\%)} & \textbf{F1(\%)} \\ \midrule
\textbf{P - Personal}& 65.71& 62.86& 66.67& 64.71\\
\textbf{C - Connection}& 57.14& 65.71& 56.10& 60.53\\
\textbf{T - Textual}& 58.57& 62.86& 57.89& 60.27\\
\textbf{S - Search}& 54.29& 62.86& 53.66& 57.89\\
\textbf{I - Interaction}& 64.29& 71.43& 62.50& 66.67\\ 
\textbf{P + C}& 68.57& 77.14& 65.85&71.05\\
\textbf{P + T}& 65.71& 62.86& 66.67&64.71\\
\textbf{P + S}& 65.71& 57.14& 68.97&62.50\\
\textbf{P + I}& 70.00& 71.43& 69.44&70.42\\
\textbf{C+T+I+S}& 72.86& 77.14& 71.05&73.97\\
\midrule
\textbf{All}& \textbf{80.00} & \textbf{74.29}& \textbf{83.87}&\textbf{78.79}\\
\bottomrule
\end{tabular}%
}
\end{table}

\section{Conclusion}
In this study, we construct the first Short-Form Vedio Addicton (SFVA) dataset and introduce \textit{EarlySD}, an innovative framework for SFVA detection. We approach SFVA detection as a node classification problem, initially employing a large language model to extract comprehensive information from user-generated content. This addresses the significant challenges of sparsity and missing values in our constructed social graph. \textit{EarlySD} is designed using a heterogeneous graph neural network to capture users' similarities and their interest preferences on social media platform. Extensive experiments conducted on Instagram data validate the effectiveness and superiority of \textit{EarlySD} across all relevant metrics.

\bibliographystyle{unsrt}
\bibliography{main}

\end{CJK}
\end{document}